\documentstyle[prl,aps]{revtex}

\begin{document}

\draft

\title{Invitation to quantum dynamical semigroups}

\author{Robert Alicki}

\address{Institute of Theoretical Physics and Astrophysics, University
of Gda\'nsk, Wita Stwosza 57, PL 80-952 Gda\'nsk, Poland}

\date{\today}
\maketitle

\begin{abstract}
The theory of quantum dynamical semigroups within the mathematically rigorous framework
of completely positive dynamical maps is reviewed. First, the axiomatic
approach which deals with phenomenological constructions and general mathematical
structures is discussed. Then basic derivation schemes of the constructive approach
including singular coupling, weak coupling and low density limits are presented in
their higly simplified versions. Two-level system coupled to a heat bath, damped harmonic
oscillator, models of decoherence, quantum Brownian particle and Bloch-Boltzmann equations 
are used as illustrations
of the general theory. Physical and mathematical limitations of the quantum open system theory,
the validity of Markovian approximation and alternative approaches are discussed also.

\end{abstract}

\pacs{ }

\section{ Introduction}

Classical theory of  systems interacting with environment employs evolution
equations which can be written in the following abstract mathematical form
$$
{d\over dt}p_t = {\cal L}p_t\ ,\ \ t\ge 0\ .
\eqno(1)
$$
Here $p_t$ is a time dependent probability distribution over the relevant
system's configuration space $\Omega$. In particular $\Omega$ can be either
a phase-space of a system , its position or momentum (velocity) space or a discrete
set in the case of coarse-grained description or a discretized model. The most
general "continuous" form of
${\cal L}$ is a sum of a differential operator (Fokker-Planck or diffusion type) and 
a collisional integral one [1]. For a discrete version we can write (1) as a Pauli
master equation
$$
{d\over dt}p_t(k) = \sum_l \bigl( a_{kl} p_t(l) - a_{lk}p_t(k)\bigr)\ .
\eqno(2)
$$
The "gain $-$ loss" structure of the RHS of (2) guarantees preservation of 
$\sum_k p_t(k)$ while positivity of the transition probabilities per unit time 
$(a_{kl}\ge 0)$ implies $p_t(k)\ge 0$. Similar mathematical consistency conditions
are satisfied for the continuous cases due to the positivity of the diffusion
constant (or positively defined diffusion matrix) and/or positivity of the scattering
cross-sections. The solution of eq.(1) is given in terms of a Markov semigroup
$T_t = \exp\{t{\cal L}\}$ satisfying the composition law
$$
T_tT_s = T_{t+s}\ ,\  t,s\geq 0
\eqno(3)
$$
and preserving positivity and normalization of the probability distribution $p_t$.   
\par
In the end of 60-ties and begining of 70-ties one needed an analogical formalism
to describe a variety of irreversible phenomena in quantum optics, solid state physics, 
magnetic resonance,
nuclear and particle physics, etc. [2,3]. The classical probability distribution 
$p_t$ should be 
replaced by the density matrix $\rho_t$ satisfying the analog of eq.(1)
$$
{d\over dt}\rho_t =  L\rho_t\ ,\ \ t\ge 0\ .
\eqno(4)
$$
Besides many heuristic attempts the mathematically sound theory of quantum dynamical
semigoups has been developed very soon [4-7]. It is convenient for the further discussion
to distinguish two approaches in this theory - the {\it axiomatic} and  the {\it constructive} 
one.
\par
The former approach, presented in the next Section, is concentrated on general mathematical
properties of eq.(4) and its solutions which are by no means trivial and lead to the theory
of comptelely positive maps and semigroups on operator algebras. The fundamental structural
theorems will be given and illustrated by several examples.   
\par
In constructive approach one tries to derive eq.(4) from first principles. Starting with
a model of a quantum system interacting with a quantum environment one obtains Markovian
master equation (4) as an approximation to the exact reduced dynamics of the open system.
The canonical models of reservoirs and basic approximation procedures are discussed in 
Section III.
\par
In the recent years quantum theory of open systems and in particular quantum dynamical 
semigroups became again a subject of intensive investigations. This is motivated by the new
experimental techniques which allow precise monitoring of decoherence and dissipation
in microscopic and mesoscopic quantum systems [8]. Possible future applications of controlled 
quantum systems in quantum information processing and quantum computing are another sources
of this revival [9]. Both mentioned above approaches are necessary to create physically correct, 
mathematically consistent and convenient for practical applications theory.
\par
The number of publications on this topic is enormous and therefore the presented list of 
references is far from being complete, often review papers, books and other contributions to
this volume are quoted instead of the original papers.

\par
\section{ Axiomatic approach}

Consider an open quantum system with the associated Hilbert space ${\cal H}$ with the time
evolution governed by the master equation (4) for its density matrix $\rho_t$. This is
the {\it Schr\"odinger\ picture} of a time evolution which will be used in this 
paper. We leave
as a simple exercise for the reader to translate all the presented results to the
{\it Heisenberg picture}. 
\par
In order to give 
a precise mathematical meaning to eq.(4) we first extent it to the whole Banach space 
${\cal T}({\cal H})$ of 
trace-class operators acting on ${\cal H}$ equipped with the norm $\|\sigma\|_1 = 
{\rm tr}(\sigma\sigma ^*)^{1/2}$. Then $L$ is a linear, perhaps unbounded, operator defined on its
domain $D(L)\subset {\cal T}({\cal H})$. The formal solution of eq.(4) is given by the
the one-parameter semigroup $\Lambda_t , t\geq 0$
$$
\Lambda_t \sigma = \sum_{n=0}^{\infty} {t^n\over n!} L^n\sigma
\eqno(5)
$$
for a bounded operator $L$ while for an unbouded one
$$
\Lambda_t \sigma = \lim_{n\to\infty}({\bf 1} - tL)^{-n}\sigma
\eqno(6)
$$
makes sense. The semigroup $\Lambda_t$ should satisfy the following conditions
$$
\Lambda_t\Lambda_s = \Lambda_{t+s}\ -\ {\rm semigroup\ property}
\eqno(7)
$$
$$
\lim_{t\to +0}\Lambda_t\sigma = \sigma \ -\ {\rm continuity}
\eqno(8)
$$
$$
{\rm for}\ \sigma \geq 0\ ,\ \Lambda_t \sigma \geq 0 \ -\ {\rm positivity}
\eqno(9)
$$
$$
{\rm for\ any}\ \sigma\in{\cal T}({\cal H})\ ,\ {\rm Tr}\ (\Lambda_t\sigma)= 
{\rm Tr}\ \sigma \ -\ {\rm trace\ preservation}\ .
\eqno(10)
$$
We shall see that the particular feature of composed quantum systems, namely the existence 
of {\it entangled states}, makes us to replace the positivity condition (9) by a stronger one
called {\it complete positivity}.

\subsection{Complete positivity}

Consider two well-separated open quantum systems with dynamics given
by two dynamical maps $\Lambda^{(1)}$ and $\Lambda^{(2)}$ respectively. Assume also
that the initial state of a joined system is determined by the entangled density matrix
$$
\rho^{(12)} \ne \sum_j p_j \rho^{(1)}_j\otimes\rho^{(2)}_j\ .
\eqno(11)
$$
One can easily find  examples (e.g. for 2 qubits) of positive trace preserving 
$\Lambda^{(1)}$ and $\Lambda^{(2)}$ and $\rho^{(12)}$ satisfying (11) such that
$\Lambda^{(1)}\otimes\Lambda^{(2)}\rho^{(12)}$ is not positively 
defined [10]. The minimal
condition which must be imposed on dynamical maps to allow meaningful tensor products
is complete positivity which means that for any $n = 1,2,3,...$, $\Lambda\otimes I_n$
is positive,where $I_n$ is an identity map acting on $n\times n$ matrices (i.e. 
trivial dynamical map on $n$-level quantum system). Completely positive maps on operator
algebras were studied already in the 50-ties and the celebrated 
Stinespring representation [11]
leads to a general form of completely positive dynamical map called often Kraus 
decomposition [12]
$$
\Lambda \rho = \sum_{\alpha}W_{\alpha}\rho W_{\alpha}^*
\eqno(12)
$$
where $W_{\alpha}$ are bounded operators on ${\cal H}$ satisfying 
$\sum_{\alpha}W_{\alpha}^*W_{\alpha} = {\bf 1}$. The decomposition (12) is highly nonunique,
in particular the sum over $\alpha$ can be replaced by an integral. If ${\cal H}$ is 
$n$-dimensional then one can always find Kraus decomposition in terms of at most 
$n^2$ terms.

\subsection{Completely positive dynamical semigroups}

From now on by a {\it quantum dynamical semigroup} (QDS) we mean a family of maps ${\Lambda_t
 , t\geq 0}$ satisfying conditions (7)-(10) with (9) strengthened to complete positivity.
The celebrated result of  Lindblad, Gorini, Kossakowski and Sudarshan [4,5] provides us 
with the most general form of a quantum Markovian master equation (MME) with a
{\it bounded generator}
$$
{d\over dt}\rho_t = -i[H,\rho_t] + \sum_j V_j\rho_t V_j^* -{1\over 2}
\bigl\{\sum_j V_j^*V_j , \rho_t\bigr\}\ .
\eqno(13)
$$
or in a slightly different but often used form
$$
{d\over dt}\rho_t = -i[H,\rho_t] + {1\over 2}\sum_j\bigl( [V_j,\rho_t V_j^*]
+ [V_j\rho_t , V_j^*]\bigr)\ .
\eqno(14)
$$
The choice of bounded operators $H=H^*$ and $V_j$ is again not unique and the sum over
$\{j\}$ can be replaced by an integral. To simplify the notation we put always $\hbar\equiv 1$
and $k_B \equiv 1$ to have the same units for energy, frequency and temperature.
\par
The non-Hamiltonian part of the generator (13) displays the "gain - loss" structure
similar to that of Pauli master equation (2) and determined by a completely positive
{\it quantum transition map}
$$
\rho\mapsto\Phi\rho = \sum_j V_j\rho V_j^*
\eqno(15)
$$
which is a sum of "pure" transition maps $\rho\mapsto V_j\rho V_j^*$ corresponding to
elementary irreversible processes [7].
\par
The following expansion for $\Lambda_t = \exp(tL)$ involving only sums (or integrals) 
and compositions of manifestly completely positive maps $\Phi $ (15) and $W_t$
$$
W_t\rho = S_t \rho S_t^*\ ,\ \ S_t=\exp\{-itH-(t/2)\sum_j V_j^*V_j\}
\eqno(16)
$$
is often useful
$$
\Lambda_t = W_t + \sum_{n=1}^{\infty} \int_0^t dt_n\int_0^{t_n}dt_{n-1}\dots\int_0^{t_2}dt_1
W_{t-t_n}\Phi W_{t_{n}-t_{n-1}}\Phi\dots W_{t_2-t_1}\Phi W_{t_1}\ .
\eqno(17)
$$
For open systems with infinite dimensional Hilbert spaces we expect that the generators
of QDS are typically unbounded. Although there exists no theorem giving the most general 
structure of $L$ in this case, the standard form (13)(14) makes sense very often [13,14]. 
Then the operators
$H$ and $V_j$ can be unbounded, in fact $V_j$ can be even operator-valued distributions
(e.g. quantum fields) and the sum over $\{j\}$ can be replaced by an integral. The typical
mathematical condition which could be imposed is the existence of the contacting semigroup
$S_t$ (16) on the Hilbert space such that the operators $S_t V_j$ became bounded and the 
expansion (17) makes sense. One of the unexpected features of the solutions to MME
with unbounded generators is that despite the formal trace preservation of (13)(14) we can have
${\rm Tr}\rho_t < {\rm Tr}\rho_0$. This phenomenon is known for classical Pauli master equations with
infinite number of states where for certain transition probabilities "the system 
can escape to infinity in finite time" [15]. 

\subsection{Examples}
We present few examples of QDS generators obtained using axiomatic construction based
on phenomenological arguments. For other interesting models of dissipative quantum systems
studied using  numerical computations also, see [16].

\subsubsection{2-level system} 

The simplest and the oldest example of QDS for a 2-level system
(qubit) can be constructed using three elementary transition maps. Assume that 
$|1>$ and $|2>$ form a basis of Hamiltonian eigenvectors and define the following operators
$$
P_k = |k><k|\ ,\  k=1,2,\ \sigma ^+ = (\sigma^-)^* = |2><1|,\ \sigma_3 = P_2 -P_1\ .
\eqno(18)
$$
The transition maps correspond to three different irreversible processes
$$
{\rm damping}\ \ \rho\mapsto \gamma_{\downarrow}\sigma^-\rho\sigma^+
\eqno(19)
$$
$$
{\rm pumping}\ \ \rho\mapsto \gamma_{\uparrow}\sigma^+\rho\sigma^-
\eqno(20)
$$
$$
{\rm pure\ decoherence}\ \ \rho\mapsto \delta_1 P_1\rho P_1 + P_2\rho P_2\ .
\eqno(21)
$$
The master equation obtained from (18)-(21) reads
$$
{d\over dt}\rho_t = -{i\omega\over 2}[\sigma_3 , \rho_t] + {\gamma_{\downarrow}\over 2}
\bigl([\sigma^- ,\rho_t\sigma^+] + [\sigma^- \rho_t ,\sigma^+]\bigr) 
 +{\gamma_{\uparrow}\over 2}
\bigl([\sigma^+ ,\rho_t\sigma^-] + [\sigma^+ \rho_t ,\sigma^-]\bigr) 
- {\delta\over 2}[\sigma_3 ,[\sigma_3 ,\rho_t]]
\eqno(22)
$$
where $\delta = (\delta_1 +\delta_2)/2$. Using the decomposition 
$$
\rho_t = p_1(t)P_1 + (1-p_1(t))P_2 + \alpha(t)\sigma^+ +{\bar\alpha}(t)\sigma^-
\eqno(23)
$$
we obtain 
$$
p_1(t) = p_1(0) e^{-(\gamma_{\uparrow}+\gamma_{\downarrow})t} +
{\gamma_{\downarrow}\over \gamma_{\downarrow} +\gamma_{\uparrow}}
\bigl(1-e^{-(\gamma_{\uparrow}+\gamma_{\downarrow})t}\bigr) 
 \ ,\  \alpha (t) = \alpha (0) \exp\{-i\omega t - (\gamma_{\downarrow}
+\gamma_{\uparrow}+\delta)t/2\}.
\eqno(24)
$$
The diagonal elements of $\rho_t$ evolve independently of the off-diagonal ones
and the former satisfy Pauli master equation with a stationary state which can be
written in  a form of the Gibbs state 
$$
\rho_{eq}= \bigl(2\cosh {\omega\over 2T}\bigr)^{-1} e^{- \omega\sigma_3/2T}\ ,\  
{\gamma_{\uparrow}\over \gamma_{\downarrow}} = e^{-\omega/T}\ . 
\eqno(25)
$$
Moreover, for any initial state $\rho_t \to\rho_{eq}$ for $t\to\infty$ and the generator 
satisfies {\it quantum detailed balance} condition. 
\par
The equation (22) written explicitly in terms of matrix elements is called {\it Bloch
equation} and is frequently used in quantum optics and magnetic resonance theory.

\subsubsection{Harmonic oscillator} 

Another well-known example is a linearly damped and pumped
harmonic oscillator with a Hamiltonian $H = \omega a^+a$ ($[a,a^+]=1$) and transition
maps describing coherent damping and pumping
$$
{\rm damping}\ \ \rho\mapsto \gamma_{\downarrow}a\rho a^+\ ,\ 
{\rm pumping}\ \ \rho\mapsto \gamma_{\uparrow} a^+\rho a \ .
\eqno(26)
$$
The master equation with an unbounded generator
$$
{d\over dt}\rho_t = -{i\omega}[a^+a , \rho_t] + {\gamma_{\downarrow}\over 2}
\bigl([a,\rho_t a^+] + [a \rho_t ,a^+]\bigr) 
 +{\gamma_{\uparrow}\over 2}
\bigl([a^+ ,\rho_t a] + [a^+ \rho_t , a]\bigr) 
\eqno(27)
$$
is exactly solvable and the solution can be presented for instance in the form
of the quantum generating function of the complex variables $z ,{\bar z}$ [7]
$$
F_t(z, {\bar z})= {\rm Tr}\bigl(\rho_t \exp\{za-{\bar z}a^+\}\bigr) = 
e^{-A(t)}F_0(z_t,{\bar z}_t)
\eqno(28)
$$
where
$$
z_t = z\exp\{-i\omega t- (\gamma_{\downarrow}-\gamma_{\uparrow})t/2\}\ ,\ 
A(t) = {|z|^2\over 4}{\gamma_{\downarrow}\over \gamma_{\downarrow} -\gamma_{\uparrow}}
\bigl(1-e^{-(\gamma_{\downarrow}-\gamma_{\uparrow})t}\bigr) \ .
\eqno(29)
$$
By differentiation of $F_t (z,{\bar z})$ with respect to $z$ or ${\bar z}$ one obtains
time dependence of all moments ${\rm Tr}(\rho_t (a^+)^n a^m)$. One can see from (28)(29) that
if $\gamma_{\downarrow}>\gamma_{\uparrow}$ then $\lim_{t\to\infty}\rho_t = \rho_{eq}$
where 
$$
\rho_{eq} = \bigl( 1-e^{-\omega/T}\bigr)\exp \bigl\{-{1\over T}\omega a^+a\bigr\}\ ,\  {\omega\over T} =
 \log {\gamma_{\downarrow}\over\gamma_{\uparrow}}
\eqno(30)
$$
One can also show that the diagonal and off-diagonal matrix elements of $\rho_t$ evolve
independently with the former satisfying detailed balance Pauli master equation.
\par
Equation (27) found numerous applications, for instance in quantum optics to describe single 
mode of radiation in cavity. This is also the simplest example of the important class
of the {\it quasi-free } QDS (see [7,10] and references therein).

\subsubsection{Decoherence of  mass center} 

Following [17] we  briefly present a  model based on QDS which can be used to analyse
environmental decoherence  which cause the transition from quantum to classical world.
Consider a mass 
center of a body described by the Hamiltonian  
$$
H = {1\over 2M}{\bf P}^2 + V({\bf X})
\eqno(31)
$$
where 
${\bf X}$ and ${\bf P}$ are the operators of mass center and total momentum  
 satisfying 
$$
[X_k,P_l] = i\delta_{kl}\ ,\ k,l= 1,2,3\ . 
\eqno(32)
$$
We assume that the interaction with  an environment can be reduced to processes of 
scattering, absorption or emission.  
A collision with a gas particle and emission, absorption 
or scattering of a photon (or other quasiparticle)
causes a momentum transfer ${\bf k}$ which changes the total
momentum. The following transformation (in the Heisenberg picture)
$$
e^{i{\bf kX}} {\bf P} e^{-i{\bf kX}} = {\bf P} + {\bf k}
\eqno(33) 
$$
describes this momentum transfer independently of the detailed microscopic
mechanism of energy redistribution. Therefore the elementary irreversible process
is given by the transition map
$$
\rho \mapsto e^{-i{\bf kX}} {\rho} e^{i{\bf kX}} \ .
\eqno(34) 
$$
Assuming statistical independence of different
momentum transfer events (called simply {\it collisions}) we obtain the following
form of the MME
$$ 
{d\over dt}\rho_t = -i [H ,\rho_t] + \int_{{\bf R}^3} d^3{\bf k}\,n({\bf k})\Bigl(e^{-i{\bf kX}}\rho_t
e^{i{\bf kX}}-\rho_t\Bigr)
\eqno(35)
$$
where $n({\bf k})$ is a density of collisions per unit time leading to the momentum 
transfer ${\bf k}$.
\par
The generator given by (35) takes into
account momentum conservation but  the average kinetic energy
grows to infinity for $t\to\infty$ and hence the process of ultimate 
relaxation to equilibrium
is not properly described. However, it is expected that in the limit of large mass 
$M$ and for slowly varying potential $V({\bf x})$ the decoherence
time is much shorter than the energy dissipation time scale. 
Therefore the master equation (35) can be considered as a good approximation for the study of 
pure decoherence in the relevant regime of large body at slowly varying potential 
and rare collisions (see Section III. D.3).
\par
\subsubsection{Bloch-Boltzmann equation}

In quantum optics the investigation of an active atom interacting with radiation and
immersed in the thermal bath of perturbers (typically , noble gas atoms) is a very important
topic [18]. To simplify the description one assumes that the spatially uniform distribution
of active atoms and strong decoherence due to collisions with perturbers justify
the use of the density operators diagonal in momentum (or velocity) representation.
Therefore, we consider a family $\rho ({\bf v})\geq 0$ of velocity-dependent positively
defined operators (matrices) acting on the Hilbert space describing the internal structure
of an active $n$-level atom. Using a standard form of the master equation (13)
one can easily derive the most general evolution equation for QDS which preserves such 
partially diagonal structure of the density matrix [19]. It reads
$$
{d\over dt}\rho_t({\bf v})= -i \sum_a h_a({\bf v})[S_a ,\rho_t({\bf v})] +
\sum_{a,b}\int_{{\bf R}^3} d^3{\bf v'}{\cal K}_{ab}({\bf v};{\bf v'}) S_a \rho_t({\bf v'})S_b^*
-{1\over 2}\sum_{a,b}\gamma_{ab}({\bf v}) \{S_a^* S_b , \rho_t({\bf v})\}
\eqno(36)
$$ 
where $\{S_a \}$ form a linear basis in the space of $n\times n$ matrices and
$$
{\rm for\ any}\ {\bf v},{\bf v'},\ \ \ {\cal K}_{ab}({\bf v};{\bf v'})\ \  {\rm is\ positively\ 
defined}\ ,\ \
\gamma_{ab}({\bf v}) = \int_{{\bf R}^3} d^3{\bf v'}{\cal K}_{ba}({\bf v};{\bf v'})\ . 
\eqno(37)
$$ 
Equation (36) provides an example of a classical-quantum description which combines
a generalisation to $n$-level case of the MME (22), called Bloch equation 
with a classical
linear Boltzmann equation for spatially uniform case. Similar type of equations found applications
in the quantum measurement theory [20].

\subsection{ Ito-Schr\"odinger equations}

In the classical theory the Fokker-Planck equation describing Brownian motion in terms
of probability distribution over the phase-space of Brownian particle is completely
equivalent to the Langevin equation (in Ornstein-Uhlenbeck formulation) for the particle's 
random trajectory. The later equation is a prototype of {\it stochastic differential
equation} which can be written either in Ito or Stratonovitch form [1]. The similar equivalent
description exists for quantum dynamical semigroups with the generator (13) in terms of
the following {\it Ito-Schr\"odinger equation}
$$
d\psi (t) = -iH\psi (t) dt - {1\over 2}\sum_j V^*_jV_j \psi(t) dt -i\sum_j V_j \psi(t) dB_j(t)\ .
\eqno(38)
$$
In (38) $B_j(t)$ denote independent and normalized Brownian motion processes and heuristically
"${dB\over dt}$" corresponds to "white noise".
The short-hand notation $df(t) = a(t)dt + b(t) dB (t)$ has a meaning of the corresponding 
integral equation 
$$
f(t_1) - f(t_2) = \int_{t_1}^{t_2} a(t) dt +  \int_{t_1}^{t_2} b(t) dB(t)
\eqno(39)
$$
where the second integral is an "nonanticipating" Ito integral given by the probabilistic limit
of the sums $\sum b(s_k) [B(s_{k+1}) - B(s_k)]$ with $ t_1 = s_1 < s_2<\dots <s_n = t_2$.
To perform computations using stochastic integrals one applies Ito rule 
$$
dB_i(t) dB_j(t) = \delta_{ij} dt
\eqno(40)
$$
and $E[dB_j(t)] = 0$ where $E[\cdot]$ denotes the average with respect to 
Brownian motions. One should remember that $dB_j(t)= B_j(t+dt)- B_j(t)$ is always statistically
independent on the solution of stochastic differential equation up to time $t$. 
Applying those rules to the density matrix of the open system defined as an average
$\rho (t) = E[|\psi (t)><\psi(t)|]$ we obtain
$$
d \rho_t = d E[|\psi (t)><\psi(t)|] = E[|d\psi (t)><\psi(t)| + |\psi (t)><d\psi(t)|
+|d\psi (t)><d\psi(t)|]
$$
$$
 = -i [H, \rho_t] dt -{1\over 2} \{\sum_j V^*_jV_j , \rho_t\}dt
+ \sum_j V_j\rho_t V_j^* dt
\eqno(41)
$$
that coincides with MME (13). 
The Ito-Schr\"odinger equation (38) for a general case was firstly proposed in [21,22]and then
derived by different authors as a useful tool for numerical simulations of quantum open 
systems [23].
For such applications it is convenient to use an equivalent nonlinear modification of (38) 
with the
normalized solutions $\phi(t) = \psi(t)/\|\psi (t)\|$. One should notice that to obtain
(41) we can replace the Brownian motions $B_j(t)$ by arbitrary stochastic 
processes $X_j(t)$ ( continuous or jump processes) with independent increments such that 
$E[dX_i(t) dX_j(t)] = \delta_{ij} dt$. 
The different processes represent unitarily equivalent representations     
of the singular reservoirs constructed in terms of "time ordered Fock spaces" [21]. 
Possible physical interpretations of these different representations are discussed in [24]. 
\par
An alternative approach in terms of "quantum noises" has been proposed in [25]
and subsequently developed into a mathematically rigorous theory of noncommutative stochastic
differential equations.

\subsection {Problems and pitfalls}

Although, the axiomatic approach allows to write down the MME (13,14)
in a mathematically consistent form, we generally need additional phenomenological insights 
which could lead to concrete form of the operators $\{V_j\}$. The first problem is the relation 
between
the dissipative part in (13,14) and the Hamiltonian of the open system. Here the demanded
structure of the stationary state and detailed balance condition are helpful [7].For example
adding a nonlinear term to the Hamiltonian of the harmonic oscillator we have to modify the
dissipative part too, in contrast to often used simplified models with linear dissipation. 
The symmetries of the system and reservoir give another restrictions on the form of generators
[26].
\par
For many interesting physical problems we have to include time dependent-external fields 
acting on the system. Very often it is done by simply adding the proper time-dependent term
to the Hamiltonian without changing the dissipative part. This can be justified, however,
for weak and quickly oscillating fields while in general case the whole Markovian approximation
breaks down [27]. Another extreme case is a slowly varying external potential which leads to a 
proper time-dependent generator which for any time reproduces, for example, return to an
equilibrium state given by a total instant Hamiltonian [28].
\par
A particular care is needed when we put two open systems $S_1,S_2$ with their "privat" reservoirs
$R_1,R_2$ into contact by adding their interaction Hamiltonian $V_{12}$.  In this case we cannot
simply add two dissipative generators obtained previously for decoupled systens $S_1,S_2$. This
is obvious for the case of two heath baths at the same temperature which should drive the 
interacting
system $S_1 +S_2$ into its Gibbs state which of course depends on the new element $V_{12}$.

\section{Constructive approach}

The goal of the constructive approach is to derive, using mathematically controlled
approximations, Markovian dynamics of the quantum system interacting with the quantum 
environment. We restrict ourselves to the derivations which respect complete positivity
of the reduced dynamics. In author's opinion the very scheme of quantum open systems approach
and particularly Markovian approximations are consistent only under well defined
conditions discussed below. 

\subsection{Basic assumptions} 

In the theory of quantum open systems we decompose the Universe into three parts ($S,O,R$).
The first one is an open system $S$ controlled by an "active" observer $O$ and interacting
with the "passive" rest 
$R$ which is called reservoir. If the state of the system $S+R$ is given by the density matrix
$\rho_{SR}$ that we asume that the reduced density matrix $\rho_S = {\rm Tr}_R \rho_{SR}$
possesses an operational meaning i.e. that the observer $O$ can prepare an arbitrary
initial state of $S$
at a certain moment $t_0 = 0$ and than measure the evolved state at the moment $t>0$. 
We often assume also that the observer can control to some extend the evolution of $S$ 
switching-on external time-dependent "fields". 
The physical structure of $O$ and the description of measurements performed na  $S$ is a subject
of quantum measurement theory and will be not discussed here. 
\par
One should notice that the above scheme contains a tacit assumption that the interaction between
$S$ and $R$ is weak. Any preparation of the initial state of $S$ or its state measurement
takes time $t_p = \hbar /\delta E$ where $\delta E$ is an energy resolution of state
preparation or measurement. For a quantum system $S$ with a discrete spectrum $\delta E$
should be smaller that the typical separation between the energy levels while for systems
with continuous spectrum (e.g. quantum Brownian particle) $t_p$ is directly related
to a chosen time-scale of observation (see Section III.D.3). Obviously, during preparation or 
measurement processes the system
$S$ should not be strongly perturbed by the environment $R$ what implies the inequality
$$
t_p << \tau_S
\eqno(42)
$$
where $\tau_S$ is a characteristic time scale of relaxation of $S$ due to the interaction 
with $R$.
\par
The first step towards mathematical description of the open system consists in writing
down the formal "bare" Hamiltonian 
$$
H_{SR} = H^0_S +\lambda \sum_k S_k\otimes R^0_k + H_R
\eqno(43)
$$
where $H^0_S$ is a bare Hamiltonian of $S$, the interaction term is proportional to
a dimensionless coupling constant $\lambda$ and contains bare reservoir's operators
$R^0_k$. For some applications we shall add a time-dependent contribution $V(t)$ to
the Hamiltonian $H^0_S$ in order to take into account external fields used by the observer $O$
to control our system $S$. 
\par
The next important assumption is the product structure of the initial state for 
$S+R$
$$
\rho_{SR}(0) = \rho\otimes \omega_R
\eqno(44)
$$
with an arbitrarily prepared initial state of $S$ and a fixed reference state of $R$.
Again this assumption can be justified if the interaction is weak in a defined below sence [29].
Namely, the system $S+R$ for times preceding the preparation at $t_0 =0$ is
expected to be at equlibrium or ground state $\omega_{SR}$. In order to produce
the initial state (44) by applying external perturbations to $S$ we must assume that
$$
\omega_{SR} \simeq \omega_S\otimes \omega_R\ .
\eqno(45)
$$
The state of $S$ for times $t>> t_p$ is then given by the reduced density matrix (we omit
subscript $S$)
$$
\rho_t = {\rm Tr}_R\bigl( U_t \rho\otimes\omega_R U_t^*\bigr) = \Lambda_t(\rho)
\eqno(46)
$$
where $U_t = \exp\{-it H_{SR}\}$ and ${\rm Tr}_R$ is a partial trace over the Hilbert space 
of $R$. 
The dynamical maps $\{\Lambda_t ; t\geq 0\}$
are by construction completely positive and trace preserving.
\par
However, the practical implementation of the above prescription leads to difficulties.
Take as an example, a model of atom interacting with electromagnetic field at zero temperature.
The corresponding Hamiltonian (43) is well-defined only if an ultraviolet
cut-off at the frequency $\omega_c$ is introduced replacing "bare" operators $R^0_k$ by
cut-off dependent regularized $R_k$. Then a simple lowest order evaluation of (46) shows
that the interaction produces a cut-off dependend Hamiltonian correction to the bare Hamiltonian
$H^0_S$. When $\omega_c\to\infty$ corrections diverge and must be cancelled by proper 
counterterms in the bare Hamiltonian (renormalization procedure). In the case of our example
it leads to mass and charge renormalization and slight modification of the Coulomb potential
which influences energy levels of the atom (Lamb-shift)[30]. A different example is an electron
immersed into a polar solid. The interaction with lattice ions modelled by the electron-phonon
coupling (with natural well-defined cut-off) completely changes the propeties of an electron
leading to the so-called polaron model. In both cases the initial bare system is strongly
modified by a "cloud" which consists of certain degrees of freedom of the reservoir [31]. Hence the 
decomposition into $S+R$ is not unique and the practical criterion should be the weakness
of the residual interaction between the "dressed" system $S'$ and the rest of the reservoir
$R'$. Practically, we describe the sytems $S',R'$ by the same Hilbert spaces as $S,R$ and
the Hamiltonian of the bath $H_R$ remains the same too. We introduce, however, a physical
(renormalized) Hamiltonian $H_S$ and modify the structure of interaction Hamiltonian introducing
suitable formfactors with cut-offs and often leaving only resonant terms ("rotating wave
approximation" in quantum optics, Lee models in particle physics) to reduce creation of the 
"cloud". We shall ilustrate these problems later on in the case of 2-level open system
(see Sections III.C.2, III.D.2).

\subsection{Models of reservoirs} 

Similarly to the classical case represented by the standard example of Brownian motion
the details of the reservoir  should not be essential for the
dynamical behaviour of an open system $S$ at least on the proper "coarse-grained" time
scale. Instead we expect that the mechanisms illustrated by limit theorems in classical
probability should work in the quantum domain also. Indeed, take a sequence of quantum
observables $A_1,A_2,...,A_N$ which are statistically independent (i.e. $<A_k A_l> =
<A_k><A_l>$) or even weakly dependent. Then, one can prove a quantum version of the Central
Limit Theorem for $N\to\infty$
$$
{1\over \sqrt N}\sum_{j=1}^N (A_j - <A_j>)\rightarrow a(f_A) + a^+(f_A)
\eqno(47)
$$
where $a(f_A), a^+(f_A)$ are "smeared" bosonic anihilation and creation operators [32,33].
The limit is understand in the sense of convergence of all correlation functions, where
for Bose fields we choose a "vacuum" as a reference state. The detailed structure of the
Fock space for bosonic field is discussed in [10,32,33] and is not relevant here.
One should think about
bosonic fields as quantum counterparts of classical Gaussian random fields and the "vacuum"
can represent any quasi-free state of bosons as for example arbitrary Gibbs state for 
noninteracting Bose gas. The index $j$ for $A_j$ can correspond both to different
parts of the reservoir and to different time instants . The limit theorem (47) reflects
the physical picture of influence from $R$ on $S$ which can be decomposed into a sum of many 
almost independent 
small contributions. Such case is usually called {\it diffusive regime} and 
due to quantum central limit theorem (47) is well described by
the harmonic oscillators bath with a weak linear coupling in the interaction
Hamiltonian
$$
H^{int}_{SR} = \lambda \sum_{k} S_k\otimes [a(f_k) + a^+(f_k)]\ .
\eqno(48)
$$
Diffusive regime does not describe all physically interesting situations. Another,
{\it low density regime} corresponds to rare events of scattering by essentially independent
"particles" from $R$ each of the scattering processes takes short time but need not to be weak.
This is similar to classical limit theorems leading to Poisson distributions and processses.
The physical model of such reservoir is a free fermionic or bosonic gas and the bilinear
interaction given by
$$
H^{int}_{SR} = \sum_{k} S_k\otimes [a^+(f_k)a(g_k)+ a^+(g_k)a(f_k)]
\eqno(49)
$$
with a small parameter being now the density of gas $\sim < a^+(f_k)a(f_k)>$.
\par
For some applications the quantum nature of the reservoir is not very important, 
like for instance in the {\it high temperature regime}. In this case one can replace
a quantum reservoir $R$ by a time dependent random Hamiltonian or in  other words
operator-valued (self-adjoit) stochastic process. A particular choice of white-noise process
leads to a stochastic Ito-Schr\"odinger equation (38) with $V_j = V_j^+$.

\subsection{ Markovian limits}

The general reduced dynamics (46) does not satisfy the semigroup composition law (7).
However, one expects that at least for a certain coarse-grained time scale, roughly determined
by $\tau_S$ (see (42)), this law is
often a very good approximation. Physically, it is true if the exact state of the system
$S+R$ given by  $U_t \rho\otimes \omega_R U_t^*$ does not differ {\it locally} from the
state $\Lambda_t(\rho)\otimes \omega_R$. Here "locality" is determined by the radius of 
interaction between $S$ and $R$. Mathematically this condition can be expressed as 
a sufficiently fast decay of the reservoirs correlation functions
$$
R_{kl}(t) = {\rm Tr}(\omega_R R_k(s+t) R_l(s))
\eqno(50)
$$
where $R_k(t) = \exp(itH_R)R_k \exp(-itH_R)$ and we assume that the reservoir's reference
state $\omega_R$ is stationary with respect to its evolution. We shall argue that those
correlation functions contain the total information about reservoirs which is relevant
for different Markovian regimes. Introducing the reservoir's
relaxation time scale $\tau_R$ we may write the standard condition for the validity
of the Markovian approximation as
$$ 
\tau_R << \tau_S\ .
\eqno(51)
$$
However, as we shall see, important examples of reservoirs may not have a natural 
decay time scale $\tau_R$ and one has to include the averaging effect of
the Hamiltonian
dynamics of $S$ in order to satisfy (51). Moreover, rigorous analysis shows  that the
Markovian approximation needs also sufficently rapid decay of higher order multitime correlation
functions which can be proved for quasi-free models of reservoirs discussed in the previous
subsection.
\par 
In the present review we are not going to reproduce involved rigorous derivations
which can be found in the literature [34-36]. Our aim is to discuss physical assumptions
behind different Markovian regimes and proper approximation schemes based on the relevant
order of perturbation which yield mathematically consistent results. For notational
simplicity we use as an illustration the model of an effective 2-level open system with 
a physical (renormalized)
Hamiltonian $H_S$ and the interaction one $H_{int}$ of the form
$$
H_S = {1\over 2}\epsilon\sigma_3 \ \ ,\ \ H_{int} = \lambda \sigma_1\otimes R
\eqno(52)
$$
where $\sigma _k , k = 1,2,3$ are Pauli matrices. We always assume that 
$$
 {\rm Tr}(\omega_R R) = 0\ .
\eqno(53)
$$
In the heuristic derivations of MME the lowest order expansion for quantum 
dynamical semigroup of the form
$$
\rho_t =  e^{-itH_S}\bigl( \rho + \sum_j \int_0^t V_j(s)\rho V_j^*(s) -
{1\over 2} \{V_j^*(s)V_j(s),\rho\} ds\bigr)e^{itH_S} + {\cal O}(t^2)
\eqno(54)
$$ 
is compared with the relevant term in the expansion for the reduced dynamics (46). 
More precisely, it is enough to
compare only the transition map term ("gain") 
$$
\rho \mapsto\sum_j \int_0^t V_j(s)\rho V_j^*(s)ds
\eqno(55)
$$ 
with the corresponding expression
obtained from (46) using different assumption concerning mainly important time scales 
in the joint system $S+R$. Any Hamiltonian-type corrections are put equal to zero according
to the renormalization procedure.

\subsubsection{Singular coupling or white noise Anzatz} 

Using second order Dyson expansion for the total dynamics of $S+R$ with the interaction
Hamiltonian treated as a perturbation we obtain the following formula for the manifestly
completely positive "transition map" which has to be compared with (55)
$$
\rho \mapsto  \int_0^t ds_1 \int_0^t ds_2 {\rm Tr}_R \bigl( H_{int} (s_1)
\rho\otimes\omega_R H_{int}(s_2)\bigr)\ .
\eqno(56)
$$
For our 2-level system we obtain
$$
\rho\mapsto\lambda^2\int_0^t ds_1 \int_0^t ds_2 R (s_2 -s_1)\Bigl( e^{i\epsilon (s_2 -s_1)}
\sigma^- \rho \sigma^+  + e^{-i\epsilon (s_2 -s_1)}\sigma^+ \rho \sigma^- 
+ e^{-i\epsilon (s_2 + s_1)}\sigma^- \rho \sigma^- 
+ e^{i\epsilon (s_2 + s_1)}\sigma^+ \rho \sigma^+ \Bigr)\ .
\eqno(57)
$$ 
The simplest approximation which produces the Markovian transition map term (55)
is based on the assumption that the correlation time of the reservoir is much
shorter than any other relevant time scale i.e.
$$
\tau_R << {\rm min} \{ \tau_S , \tau_H \}
\eqno(58)
$$
where $\tau_H$ denotes Heisenberg time scale for $S$ related to the average energy level
spacing $\Delta E$ by $\tau_H = (\Delta E)^{-1}$. Under the condition (58) we can use the
{\it white-noise Anzatz}
$$
R(t) \simeq {\hat R}(0)\delta (t)\ ,\  {\hat R}(\omega)= \int_{-\infty}^{\infty}
R(t) e^{it\omega}dt
\eqno(59)
$$
which leads to the QDS-type of the transition map corresponding to the following MME 
$$
{d\over dt} \rho = -{i\over 2}\epsilon [\sigma_3 ,\rho] - {\lambda^2\over 2}{\hat R}
(0)[\sigma_1 ,[\sigma_1 ,\rho ]]\ .
\eqno(60)
$$
For a general case the substitution of the type (59) leads to MME (13,14) with $V_j =V_j^*$
which can be rewritten in a {\it double commutator form}. Moreover, the dissipative part does not
depend on the Hamiltonian one and we can even add a  time-dependent term $V(t)$ to obtain
$$
{d\over dt}\rho_t = -i[(H_S + V(t)),\rho_t] -{1\over 2} \sum_j [V_j ,[V_j , \rho_t]]\ .
\eqno(61)
$$
The dynamics generated by (61) is very special because the dynamical maps $\Lambda_t$ are
{\it bistochastic}, i.e. $\Lambda_t {\bf 1} = {\bf 1}$. It means, depending on the physical 
context, that the {\it infinite temperature} or {\it microcanonical} state is preserved
and the H-theorem holds i.e. for any solution of (61)[37]
$$
S(\rho_t) \geq S(\rho_s)\ , {\rm for}\ t\geq s\geq 0\ ,\ S(\rho) = -{\rm Tr}\rho\ln\rho\ .
\eqno(62)
$$
There are several instances where the equations of type (61) are useful and provide reasonable
approximations to the exact dynamics of the open system.
These are the cases where the Hamiltonian
self-evolution due to $H_S+V(t)$ is very slow or $H_S +V(t)$ commutes with $S_k$ in (43)and 
therefore the pure decoherence processes dominate over the energy redistribution. 
Another case is the high
temperature limit where $T >> E_{max}$ and $E_{max}$ is a maximal difference of energy levels
of the system. The white noise Anzatz can be obtain from the Hamiltonian models by limit
procedure called {\it singular coupling limit} which leads to unphysical limiting reservoirs
with unbounded from below Hamiltonians [35].
\par
One should remember, however, that the lack of memory expressed by the $\delta$-like
correlations in (59) contradicts the quantum nature of the reservoir (see Section III.D.1)
and the noise governing the dissipative part of (61) is essentially classical.
Hence, in author's opinion, equations of the type (61) are too rough to describe properly 
the control
of decoherence and dissipation in the context of quantum computations and error correction
schemes. The same is true for their discrete-time versions which often appear in the literature
on this topic [9].

\subsubsection{Weak coupling limit} 

Another approximation scheme takes into account the interplay between the Hamiltonian dynamics
of $S$ governed by $H_S$ and the coupling to $R$. It allows to describe, for instance,
the equilibration process leading in the case of a heat bath to a final Gibbs state of $S$.
We assume the following relations between the relevant time scales
$$
\tau_H << \tau_S\ ,\ , \tau_R << \tau_S\ .
\eqno(63) 
$$
As a consequence of (63) in the example (57) the last two non-resonant terms can be omitted
and the obtained transition map corresponds to the following MME
$$
{d\over dt} \rho_t = -{i\over 2}\epsilon [\sigma_3 ,\rho_t] - {\lambda^2\over 2}
\Bigl({\hat R}(\epsilon)\bigl([\sigma^- ,\rho_t \sigma^+] +[\sigma^- \rho_t ,\sigma^+]\bigr) 
+ {\hat R}(-\epsilon)\bigl([\sigma^+ ,\rho_t \sigma^-] +[\sigma^+ \rho_t ,\sigma^-]\bigr)
\Bigr)\ . 
\eqno(64)
$$
which is a special case of (22) with $\delta = 0$. The similar procedure applied to 
a harmonic oscilator linearly coupled to an environment yields MME (27)
\par
For a general case (43) the MME obtained under the conditions (63) reads
$$
{d\over dt} \rho_t = -i[H_S ,\rho_t] + {\lambda^2\over 2}
\sum _{\omega; k,l}{\hat R}_{kl}(\omega)\bigl([S_k (\omega),\rho_t S^*_l(\omega)]
+[S_k (\omega)\rho_t , S^*_l(\omega)]\bigr)
\eqno(65)
$$
where
$$ 
S_k (t) = e^{itH_S} S_k e^{-itH_S} = \sum _{\omega} S_k (\omega)e^{-i\omega t}
\eqno(66)
$$
and
$$
{\hat R}_{kl}(\omega)= \int_{-\infty}^{\infty}
R_{kl}(t) e^{-it\omega}dt\ .
\eqno(67)
$$
Obviously, as for any eigenfrequency $\omega$ the matrix $[{\hat R}_{kl}(\omega)]$ is positively defined
we can always rewrite (65) in a diagonal form (13,14).
\par
The approximation procedure leading to MME (65) ( in the interaction picture) can be made
mathematically rigorous introducing the concept of weak coupling or van Hove limit
applied to the reduced dynamics in the interaction picture which consists of 
using the rescaled time
$\tau = \lambda^2 t$ for $\lambda\to 0$. The details can be found in [34] see also
the related idea of stochastic limit in [33].
\par
MME of the type (65) possess several interesting properties. The dissipative part
of the generator commutes with the Hamiltonian one and the diagonal elements of $\rho_t$
in the energy representation evolve independently of the off-diagonal ones. The corresponding
transition probabilities which appear in the Pauli master equation for diagonal elements
are exactly equal to those calculated using the Fermi Golden Rule [38]. In the frequently used
case of the reservoir being at thermal equilibrium  (heat bath) the KMS relation of the form
$$
{\hat R}_{kl} (-\omega) = e^{-\beta\omega} {\hat R}_{lk}(\omega)\ ,\ \beta = T^{-1}
\eqno(68)
$$
is valid and implies that that the Gibbs state $\rho_{eq}= Z^{-1} \exp\{-\beta H_S\}$
is an invariant state for MME (65). If for all $k,\omega $,
$[S_k (\omega), X]=0$ implies $X= c{\bf 1}$ the Gibbs state is ergodic i.e. for $t\to\infty$ 
and any initial state the solution of (65) $\rho_t \to\rho_{eq}$. Moreover, the Pauli master
equation for the diagonal elements of $\rho_t$ satisfies detailed balance condition. The part
of the generator in (65) which is given by terms with $\omega = 0$ does not influence
the diagonal elements of $\rho_t$ and describes pure decoherence which is not accompanied by
the energy change.
\par
One should notice the important differences between the regimes characterized by inequalities
(58) and (63) respectively. The former allows white-noise Anzatz and hence the strictly
Markovian approximation entirely due to the memoryless reservoir. In the later case
Markovian approximation is valid for the time evolution averaged over many periods of
Hamiltonian evolution of the open system. It follows that the dissipative part of the
semigroup generator strongly depends of the Hamiltonian $H_S$ and therefore we cannot
freely add a time dependent part $V(t)$ to $H_S$.

\subsubsection {Weak coupling and thermodynamics}

We have already seen that the MME (65) obtained using weak coupling limit satisfies
the {\it 0-law of thermodynamics}, namely the heat bath drives an open system $S$ to its
equlibrium state at the same temperature. Introducing a time dependent Hamiltonian
$H_S(t)$ which varies on the time-scale much longer than all other time-scales in
(63) we can apply the same method used to derive MME (65). As a result of such
combined weak-coupling and adiabatic approximation we obtain an inhomogeneous in time
MME of the form
$$
{d\over dt}\rho_t = -i[H_S(t),\rho_t] + L_D(t)\rho_t \equiv L(t)\rho_t\ ,\ \ t\ge 0\ .
\eqno(69)
$$
where the dissipative part $L_D(t)$ is computed according to the formulas (65,67) with
the Hamiltonian $H_S$ replaced by its temporal value $H_S(t)$. We can now formulate
the {\it I-st law of thermodynamics} as
$$
{d\over dt} E(t) =  {d\over dt} W(t) + {d\over dt} Q(t) 
\eqno(70)
$$
where $E(t) = {\rm Tr}(\rho_t H_S(t))$ is an internal energy of $S$, 
$W(t) = \int_0^t {\rm Tr}(\rho_s {d\over ds} H_S(s))ds$ is the work performed on $S$ by external
forces and $Q(t)= \int_0^t {\rm Tr}\bigl[({d\over ds}\rho_s) H_S(s)\bigr]ds$ is the heat 
supplied to $S$ by $R$ [28,39].
\par
In order to illustrate the second law of thermodynamics we use the {\it relative entropy}
for a pair of density matrices $\rho,\sigma$
$$
S(\rho |\sigma ) = {\rm Tr}\bigl(\rho\ln\rho - \rho\ln\sigma\bigr)
\eqno(71)
$$ 
and the following inequality
valid for any trace preserving completely positive map $\Lambda$ [37]
$$
S(\rho |\sigma )\geq S(\Lambda\rho |\Lambda\sigma )\ . 
\eqno(72)
$$ 
For the time evolution governed by the generator $L(t)$ (eq.(69)) which posseses
a temporal stationary state
$$
L(t)\rho_{eq}(t) = 0\ ,\ \rho_{eq}(t) = Z^{-1} \exp\{-\beta H_S(t)\}
\eqno(73)
$$
the inequality (72) implies the following form of the second law of thermodynamics
for open systems
$$
{d\over dt} S(\rho_t) = \sigma[\rho_t] + {1\over T}{dQ\over dt}\ .
\eqno(74)
$$
where $\sigma[\rho_t]\geq 0$ is an entropy production and the second term describes
the entropy exchande with the heat baths. The formula (74) can be easily generalized
to the case of open system coupled to several heath baths at different temperatures
[28,40].
\par
The presented derivation of the three fundamental laws of thermodynamics has been possible
within the assumptions of weak coupling between $S$ to $R$ and the adiabatic change of
external forces. Beyond this regime, even the unique definitions
of the fundamental thermodynamic notions are not obvious. One should mention that the deviations
from the thermodynamical behaviour for strongly coupled quantum systems attracted, recently,
attention of the scientific community [41].

\subsubsection{Low density limit} 

Another situation which justifies the use of MME is an open system $S$ with the discrete
spectrum Hamiltonian $H_S= \sum_k \epsilon_k |k><k|$ interacting with a dilute gas. 
The rigorous results can be found
in [33,36] while here we present a heuristic approach. Consider first an abstract 
formulation of the {\it impact approximation} for the scattering problem given by the 
Hamiltonian  $H = H_0 + V$. Starting with the identity
$$
U_t = e^{-itH} = e^{-itH_0}\Bigl({\bf 1} + \int_0^t e^{isH_0} V e^{-isH}ds\Bigr) 
= e^{-itH_0}\Bigl({\bf 1} + \int_0^t e^{isH_0} V \bigl[e^{-isH}e^{isH_0}\bigr]e^{-isH_0}ds\Bigr) 
\eqno(75) 
$$
we replace $e^{-isH}e^{isH_0}$ by $\Omega_+ = \lim_{s\to\infty}e^{-isH}e^{isH_0}$ what
makes sense for $t>> \tau_{coll}$ where $\tau_{coll}$ is a typical collision time for
the discussed model. Hence we have
$$
U_t \approx  e^{-itH_0}\Bigl({\bf 1} + \int_0^t e^{isH_0} V \Omega_+ e^{-isH_0}ds\Bigr) 
\eqno(76) 
$$
and instead of the transition map (56) we obtain its low density counterpart
$$
\rho \mapsto  \int_0^t ds_1 \int_0^t ds_2 {\rm Tr}_R \bigl( T (s_1)
\rho\otimes\omega_R T(s_2)\bigr)\ .
\eqno(77)
$$
where $T(s) = e^{isH_0}T e^{-isH_0}$ and $T= V\Omega_+$.
In the next step we obtain a transition map similar to that derived in the weak coupling
limit by averaging over the Hamiltonian evolution of $S$ and eliminating the non-resonant
oscillating terms. Take a single particle of the bath 
being in a state described by the density matrix
$$
\omega_R = {\ell}^{-3}\int_{{\bf R}^3} d^3{\bf p}\,G({\bf p})|{\bf p}><{\bf p}|\ ,\ 
<{\bf p}|{\bf p}'> = \delta^3({\bf p}-{\bf p}')    
\eqno(78)
$$
normalized in a cube ${\ell}^3$ and diagonal in the momentum representation.    
Putting $H = H_S +\int_{{\bf R}^3} d^3{\bf p}\,E_{\bf p}|{\bf p}><{\bf p}|$ we obtain
from (77) and for $t>>\tau_{coll},\tau_H$ the final form of the transition map which 
commutes with Hamiltonian evolution
$$
\rho\mapsto t{\ell}^{-3}\sum_{\omega}\int_{{\bf R}^3} d^3{\bf p} \int_{{\bf R}^3} d^3{\bf p}\, 
G({\bf p}) \pi\delta (E_{\bf p'} - E_{\bf p} +\omega) T_{\omega}({\bf p},{\bf p'})\rho
T^*_{\omega}({\bf p},{\bf p'})
\eqno(79)
$$
where
$$
T_{\omega}({\bf p},{\bf p'}) = \sum_{\epsilon_k -\epsilon_l = \omega}
<k,{\bf p'}|T |l,{\bf p}> |k><l|\ .
\eqno(80)
$$
The transition map (79) corresponds to a single-particle scattering during the time interval $t$.
Because in the volume ${\ell}^3$ we have $N$ particles the low density approximation
implies additive effect producing the factor $\nu = N/{\ell}^3$ (instead 
of ${\ell}^{-3}$) which remains finite in the thermodynamical limit. 
Therefore, the final form of the MME valid under the conditions
$$
\tau_H << \tau_S\ ,\ , \tau_{coll} << \tau_S\ .
\eqno(81) 
$$
is the following
$$
{d\over dt} \rho_t = -i[H_S ,\rho_t] +\nu\pi\sum_{\omega}\int_{{\bf R}^6} d^3{\bf p}d^3{\bf p'} 
\,G({\bf p}) \delta (E_{\bf p'} - E_{\bf p} +\omega)\Bigl([T_{\omega}({\bf p},{\bf p'}),\rho_t
T^*_{\omega}({\bf p},{\bf p'})] +[T_{\omega}({\bf p},{\bf p'})\rho_t ,
T^*_{\omega}({\bf p},{\bf p'})]\Bigr) 
\eqno(82)
$$
Similarly to the weak coupling limit the rigorous derivation of MME (82) involves the limit
procedure $\nu\to 0$ for the interaction picture version of the reduced dynamics with the 
rescaled time $\tau = \nu t$ [36].
\par 
It is not difficult to show that the MME (82) has similar properties to MME (65). Namely, 
the diagonal elements of $\rho_t$ evolve independently of the off-diagonal ones,
for the equilibrium momentum distribution of scatterers  $G({\bf p})\sim \exp(-\beta E_{\bf p})$
the Gibbs state $\rho_{eq}\sim \exp(-\beta H_S)$ is stationary and under natural conditions
ergodic. The discussion of the Section III.C.3 can be applied also for inhomogeneous in time
versions of MME (82). 
 
\subsubsection{ Application to Bloch-Boltzmann equation}

The MME (82) can be a starting point for the heuristic derivation of the Bloch-Boltzmann 
equation (36)[19]. The active atom with finite number of internal levels can be put into a finite
box to take an advantage of the dicrete spectrum of its momentum and kinetic energy operators.
To such a discretized system of an active atom we can apply the derivation of the previous section.
As an initial state we choose a quasi-diagonal density matrix
$$
\rho = \sum_{\bf v} \rho ({\bf v}) |{\bf v}><{\bf v}|
\eqno(83)
$$
with respect to the orthonormal velocity basis $\{|{\bf v}>\}$. Under some reasonable conditions
the quasi-diagonal density matrices remains quasi-diagonal during the time evolution and finally
we can go with the size of the box to infinite to obtain continuous velocity spectrum.
The final result of this procedure which involves also the transition to the center of motion
reference frame is the Bloch-Boltzmann equation (36)(37) with (we put here $\hbar$ for readers
convenience)
$$
H_S = \sum_{j=1}^n \epsilon_j |j><j|\ ,\ [H_S , S_a] = \omega_a S_a\ ,\ a=1,2,...,n^2
\eqno(84)
$$
$$
{\cal K}_{ab}({\bf v};{\bf v'}) = (2\pi)^4\hbar^2 \mu^{-3} N_p \delta_{\omega_a,\omega_b}
\int_{{\bf R}^3}d^3 {\bf v}_r'\int_{{\bf R}^3}d^3 {\bf v}_r
\delta^3 \bigl({\bf v} - {\bf v'} -{\mu\over m}({\bf v}_r - {\bf v}_r')\bigr)
\delta\Bigl({\mu\over 2}({\bf v}^2_r -{{\bf v}_r'}^2) +\hbar \omega_a\Bigr)
$$
$$
\times W({\bf v'} -{\bf v}_r') T_a({\bf v}_r,{\bf v}_r')\overline{T_b({\bf v}_r,{\bf v}_r')}
\eqno(85)
$$
where ${\bf v}_r, {\bf v}_r'$ denote relative velocities, $W_p({\bf v})$ is the equilibrium
velocity distributions of perturbers, $m$ is a mass of an active atom, $\mu$ is a reduced 
mass of
an "active atom - perturber" system, $N_p$ is a density of perturbers and the functions
$ T_a(.,.)$ are related to the $T$-matrix calculated in the center of motion reference frame by
the following expression
$$
\sum_a T_a({\bf v}_r,{\bf v}_r') S_a = \sum_{j,j'=1}^n <{\bf v}_r,j|T|{\bf v}_r', j'>
|j><j'|\ .
\eqno(86)
$$
Due to the presence of Kronecker's delta $\delta_{\omega_a,\omega_b}$ in (85) the matrix
${\cal K}_{ab}({\bf v};{\bf v'})$ is positively defined as demanded by the conditions (37).
One should mention that the obtained  Bloch-Boltzmann differs from the existing ones which
generally do not preserve positivity of $\rho({\bf v})$ [18].

\subsection{Problems and pitfalls}

The derivations of the MME presented in the previous sections are based on certain
assumption concerning the decay of correlations in the environment, separation of
energy levels of the open systems and the magnitude of the coupling constant or the
density of perturbers. These conditions are formulated in terms of relations
between the different time scales (42), (51),(63), (81). Now, we discuss  briefly
some of them and refer to the important physical situations beyond the presented
scheme stressing the related difficulties and misconceptions. 

\subsubsection{Memory effects}
The most demanding condition for the lack of memory of the reservoir is an inequality
(58) which allows the white noise Anzatz $R_{kl}(t) \simeq {\hat R}_{kl}(0)\delta(t)$.
It means that the spectral density matrix ${\hat R}_{kl}(\omega)$ is weakly dependent on
$\omega$ in the relevant energy region. Such an assumption essentially contradicts the
KMS condition (68) what is the source of the so-called {\it thermal quantum memory}
characterized by the time scale $\tau_T = T^{-1}$.
\par
Another difficulty with the assumption of memoryless reservoir can be illustrated
by a model of 2-level atom coupled to the quantum electromagnetic field at the vacuum
state ($T=0$). Puting in eq.(52) the standard "dipol$\times$ electric field" interaction
one obtains
$$
R(t)\sim {1\over (t+ i\omega_c^{-1})^4}\  ,\ \ \ {\hat R}(\omega)\sim 
\omega^3 e^{-\omega/\omega_c} 
\eqno(87)
$$
where $\omega_c$ is a cut-off frequency assumed to be larger than any energy scale relevant
for this model. The decay of reservoir's correlations is powerlike and does not possess any
natural time scale. Moreover, the correlation function is singular at the origin and 
${\hat R}(0) = 0$ for the
removed cut-off $(\omega_c\to\infty)$. Fortunately, in the weak coupling regime, the Markovian
behaviour can be restored on the coarse-grained time scale determined by $\tau_S$ - spontaneous
emission time. This is due to the averaging effect of fast Hamiltonian oscillations which
allow to replace $R(t)$ by $R(t)e^{i\epsilon t}$ - the function which effectively 
acts like ${\hat R}(\epsilon)\delta (t)$ under the integral in (57). The situation is different
when $H_S$ is replaced by a time-dependent Hamiltonian and /or collective effect for
multi-atomic systems are relevant. Namely, that fast variations of the Hamiltonian introduce
high frequency contributions increasing the decay rates (see eq.(87)) and even a system
of two atoms
possesses degenerated energy levels for which averaging effect described above does not 
apply. All that implies serious limitations on the use of Markovian approximation in the
context of controled quantum open systems, the problem which is crucial for quantum information 
processing [42].

\subsubsection{Decoherence vs. dissipation}

The dynamics of an isolated quantum system is governed by its Hamiltonian $H_S$ and is characterized
by two fundamental features : initial pure states remain pure , the average energy is a constant
of motion. On the contrary, for an open quantum system $S$ interacting with a quantum environment $R$
which starts its joint evolution from the product state, the entangled states of $S+R$ are 
developed
in the course of time what lead to the appereance of reduced mixed states of $S$ and the energy 
exchange 
between $S$ and $R$. The first phenomenon is called {\it decoherence} and the second one 
{\it dissipation}. Although decoherence and dissipation are usually present at the same time,
model calculations show that for large quantum systems approaching the border between 
quantum and classical worlds decoherence acts on a much faster time scale than dissipation 
[43,44,17].   
As decoherence seems to be a more important agent in the context of quantum measurement
theory and quantum information processing [9] it is convenient to discuss models describing 
{\it pure
decoherence} often called {\it dephasing} which is not accompanied by the energy exchange.
\par
The pure decoherence is described by the models with Hamiltonians (43) satisfying the 
condition $[H^0_S , S_k] = 0$ what implies the same condition for the renormalized
Hamiltonian $[H_S , S_k]=0$ for all $k$. In Markovian approximations the decoherence rates
are proportional to ${\hat R}_{kl}(0)$ (see (59,60) and (65-67)) which is typically
zero for the systems linearly coupled to bosonic fields . This is the case for 
electromagnetic interaction (87), the same result holds for linear coupling to phonons.
In a general case one can formulate the following "no-go theorem" for pure decoherence:
\par
{\it For quantum open systems linearly coupled to bosonic reservoirs decoherence 
is always accompanied by dissipation.}
\par
The "physical proof" is rather simple. Any irreversible decoherence must be related to an 
irreversible
change in the environment. In the case of linear coupling to bosonic field (48) this change 
can be realized only by emission or absorption of a boson - the process which changes energy
of an open system as well. Obviously, in the case of scattering process governed by the 
bilinear 
interaction Hamiltonian (49) we can alter other quantum numbers of the environment's state
(e.g. its momentum) keeping the energy of $S$ conserved (elastic scattering).
\par
One can easily find in the literature the models of dephasing based on the linear coupling
to bosonic field  which are essentially  variations of the so-called Caldeira-Leggett
model [45-47]. To explain this apparent contradiction with our "no-go theorem" consider the simplest
version of the {\it spin - boson model} defined by the Hamiltonian (here $H_S\equiv 0$)
$$
H_{SR} = \lambda\sigma_3\otimes \int_0^\infty d\omega [f(\omega)a(\omega)+ {\bar f}
(\omega) a^+(\omega)]
+ \int_0^{\infty} d\omega\,\omega\,a^+(\omega)a(\omega)
\eqno(88)
$$ 
acting on the Hilbert space 
$$
{\cal H}_{SR} = {\bf C}^2\otimes {\cal F}_B \bigl(L^2[0,\infty)\bigr))
\equiv {\cal F}_B\bigl(L^2[0,\infty)\bigr))\oplus {\cal F}_B\bigl(L^2[0,\infty)\bigr)
\eqno(89)
$$
where ${\cal F}_B \bigl(L^2[0,\infty)\bigr))$ is a bosonic Fock space over a single-particle
Hilbert space $L^2[0,\infty)$ and $[a(\omega), a^+(\omega')] = \delta (\omega -\omega')$. 
The unitary Weyl operator $U_g$ acting on fields operators as
$$
U_g a(\omega)U_g^* = a(\omega) + g(\omega) 
\eqno(90)
$$
exists if and only if $g\in L^2[0,\infty)$. Puting 
$$
g(\omega)=\lambda \omega^{-1} f(\omega)
\eqno(91)
$$ 
we can diagonalize $H_{SR}$ (88)
$$
{\bf U}_g H_{SR}{\bf U}_g^* = \int_0^{\infty} 
d\omega\,\omega\,a^+(\omega)a(\omega) +  {\rm const.}
\eqno(92)
$$
where
$$
{\bf U}_g =\pmatrix{U_g & 0      \cr
                     0  & U_{-g} \cr}\ .
\eqno(93)
$$
The ground states subspace of the diagonalized Hamiltonian 
(92) is spanned by the vectors $|1>\otimes |\Omega> ,|2>\otimes|\Omega>$ and therefore the 
corresponding degenerated ground states of $H_{SR}$ are given by
$$
|1>\otimes |\phi[-g]>\ ,\ |1>\otimes |\phi[g]>\ ,{\rm where}\  |\phi[g]> = U_g|\Omega>\ ,\ 
g(\omega) = \lambda \omega^{-1}f(\omega)\ .
\eqno(94)
$$ 
The vectors $|\phi[\pm g]>$ are coherent states in ${\cal F}_B\bigl(L^2[0,\infty)\bigr)$
and
$$
<\phi[-g],\phi[g]> = \exp\{-2\|g\|^2\}\ .
\eqno(95)
$$
Assume now that we would like to describe dephasing of our spin using the model given by(88).
Then we should have $ {\hat R}(0) = \lambda^2 |f(0)|^2 > 0$ what implies due to (91) that
the function $g(\omega)$ is not square integrable i.e. $\|g\| = \infty$. The same divergence
appears for any "ohmic" or "subohmic" coupling $|f(\omega)|^2\sim \omega ^s$ around 0
with $ 0\leq s\leq 1$. 
It follows 
that the diagonalizing transformation of a bosonic field (90)(91) cannot be implemented
by the unitary operator on the Fock space and therefore the formal expression (88) does not
define a meaningful bounded from above Hamiltonian. It means that for an arbitrary coupling
constant $\lambda$ the model given by (88) is either nonphysical or cannot describe exponential 
(Markovian) dephasing. In a less mathematical language the large value of $\|g\|$ due either
to a large coupling constant or to a large integral $\int_0^{\infty}d\omega\, \omega^{-2}
|f(\omega)|^2$ makes the coherent states $|\phi[\pm g]>$ corresponding to a "cloud"
and the vacuum $|\Omega>$ almost 
orthogonal. As a consequence the standard choice of the initial state as a product state
$\rho\otimes |\Omega><\Omega|$ is inappropriate and imposible to prepare. The proper initial
state should have a support spanned by the "dressed" ground states (94). The computed
lost of coherence for the former choice of the initial state is therefore unphysical and
describes the spurious  process of a "cloud formation". Again the problem
of a proper decomposition of the total system into  open system and environment such
that the effective interaction between them is weak, is crucial for the physical interpretation
of the obtained results.
\par
One should mention that from the mathematical point of view the above example illustrates 
the subtle problem of
nonequivalent representations of canonical commutation relations for systems with infinite
number of degrees of freedom [48].

\subsubsection{Open systems with continuous spectrum}

In the derivations of MME based on the weak coupling or low density limits the discretness 
of $H_S$ spectrum plays a crucial role. The averaging over Hamiltonian oscilations
justifies the Markovian approximation on the coarse-grained time scale and the canceling
of non-resonant terms is a necessary condition to preserve complete positivity of the QDS.
This interplay between the self-evolution of $S$ and the interaction with $R$
produces the desired properties of QDS like for instance the relaxation to a proper equilibrium
state for $R$ being a heat bath.
\par
However, there are important examples of open systems with continuous spectrum of $H_S$,
the most studied is a quantum Brownian particle. For simplicity, we discuss first the case 
of Brownian motion in one-dimensional space. The most frequently used MME for this 
case is the so-called Caldeira-Leggett equation [47] of the form 
$$
{d\over dt}\rho_t = -i[H_S ,\rho_t] -i\gamma [X, \{P,\rho_t\}] -2M\gamma T [X,[X,\rho_t]]
\eqno(96)
$$
where $X, P$ are position and momentum operators, $H_S = P^2/2M +V(X)$, $\gamma$ is a 
{\it friction 
constant} and $M$- mass of the Brownian particle. V(X) is a generic potential which can produce
both continuous and discrete parts of the energy spectrum.
\par
The MME (96) possesses the following well-known drawbacks:
\par
1) the solution of (96) does not preserve positivity of the density matrix,
\par
2) the Gibbs state (perhaps unnormalized) $\sim \exp\{-H_S/T\}$ is not its stationary
state.
\par
The first drawback can be cured by adding the term $-\gamma (8MT)^{-1} [P,[P,\rho_t]]$
which allows to write the corrected MME in a standard form (13,14). However the dissipative
part of this new generator corresponds to a damped harmonic oscillator one (27)
(with $\gamma_{\uparrow}=0$) and 
the Hamiltonian part gains the correction $\sim (XP +PX)$ with a rather unclear physical 
interpretation.
\par
In author's opinion the above difficulties have its source in the underlying Hamiltonian model
for $S+R$ system, the so-called Caldeira-Leggett model, which is essentially a 
non-zero temperature version of the model 
(88) with $\sigma_3$ replaced by the operator $X$ and the "forbidden" ohmic choice of 
$|f(\omega)|^2\sim 
\omega$ around zero. In other words according to the "no-go theorem" the equation (96) which
in the extreme heavy particle limit ($M\to\infty$)describes pure decoherence cannot 
be derived from
a physically admissible model with linear coupling to the bosonic reservoir. 
\par
It seems, that for a continuous spectrum of $H_S$ the lack of a natural time scale provided by 
the Hamiltonian evolution
makes impossible to find a single MME which accurately describes all relevant stages of the 
evolution of $S$ [49]. We do not mean here 
the well-known and expected deviations from the Markovian (exponential) behaviour for very 
short times typically $\sim \omega_c ^{-1}$
($\omega_c$ - cut-off frequency) and very long ones (due to the boundness from below
of the Hamiltonian of $R$). In the simplest case of a Brownian particle in the free space
described by the center of mass position ${\bf X}$ and the total momentum ${\bf P}$
the situation can be summarized in a following way.
\par
The proper model of an environment should involve interactions of the type (49) describing
scattering process with particles of the medium. Beside the direct collisions with atoms,
molecules, photons etc. the other "bilinear" processes are possible within this model 
like for instance absorption of a foton followed by the excitation of the internal degrees of
freedom of the Brownian particle and the time-reversed process [17].
\par
The following different but approximatively Markovian stages of the evolution can be singled 
out:
\par
1) {\it Pure decoherence stage}, when the decay of macroscopically distinguishable 
quantum superpositions into mixed states dominates over energy
thermalization. For the low density medium and/or small Brownian particle this stage is 
well-described
by the eq.(35) on the time scale determined by $\tau_{dec} = \bigl(\int d^3{\bf k}\, 
n({\bf k})\bigr)^{-1}$. For dense media or/and large particle $\tau_{dec}$ can be comparable
or shorter than the already mentioned time scale $\omega_c^{-1}$ what demands a different
theoretical treatement [50].
\par
2) {\it Thermalization stage}, when the density matrix of the Brownian particle written in
momentum representation is close to diagonal i.e. $\rho_t({\bf p},{\bf p'})$ differs essentially
from zero for $|{\bf p}-{\bf p'}|\leq\sqrt{2MT}$.   The detailed analysis of this
regime and the manifest standard form of MME which depends on the so-called dynamic structure 
factor characterizing the environment are presented in [51].
\par
Another completely different situation where the continuous spectrum of $H_S$ appears is
the theory of many-body open systems. An ensemble of interacting quantum spins coupled
to an infinite heat bath and described in the thermodynamic limit is a perfect example.
The mathematical formalism used in this context to construct proper MME, difficulties
and partial solutions of the problems are discussed in [52].

\acknowledgments
The work is supported by the Gda\'nsk University Grant BW-5400-5-0234-2.


\begin{references}

\bibitem {[1]}
C.W. Gardiner,
{\it Handbook of stochastic methods},
(Springer, Berlin, 1983).


\bibitem{[2]}
G.S Agarwal ,
{\it Quantum Optics. Quantum Statistical Theories of Spontaneous
  Emission and their Relation to Other Approaches}, (Springer, Berlin 1971).

\bibitem{[3]}
F. Haake,
{\it Statistical treatment of open systems by generalized master
  equations},
(Springer, Berlin, 1973).


\bibitem{[4]}
V. Gorini, A. Kossakowski and E.C.G Sudarshan, 
{\it J. Math. Phys.} {\bf 17}, 821 (1976).


\bibitem{[5]}
G. Lindblad,
{\it Commun. Math. Phys.}, {\bf 48}, 119 (1976).

\bibitem{[6]}
E.B. Davies, {\it Quantum Theory of Open Systems},
(Academic Press, London 1976).


\bibitem{[7]}
R. Alicki and K. Lendi,
{\it Quantum Dynamical Semigroups and Applications},
(Springer, Berlin, 1987).

\bibitem{[8]}
Ph. Blanchard et.al.(eds), 
{\it Decoherence: Theoretical, Experimental and Conceptual Problems},
(Springer, Berlin, 2000).


\bibitem{[9]} 
M.A Nielsen and I.L. Chuang, {\it Quantum Computation and Quantum Information}, 
(Cambridge University Press, Cambridge, 2000).

\bibitem{[10]}
R. Alicki and M. Fannes,{\it Quantum Dynamical Systems},
(Oxford University Press, Oxford, 2000).


\bibitem{[11]}
W.F. Stinespring, {\it Proc. Am. Math. Soc.}, {\bf 6}, 211 (1955).

\bibitem{[12]}
K. Kraus, 
{\it Annals of Physics}, {\bf 64}, 311 (1971).

\bibitem{[13]}
E.B. Davies, {\it Rep. Math. Phys.}, {\bf 11}, 169 (1977).

\bibitem{[14]}
A.M. Chebotarev and F. Fagnola, 
{\it J. Funct. Anal.}, {\bf 153}, 382 (1998).

\bibitem{[15]}
W. Feller, {\em An introduction to probability theory and its applications}
 Vol.2,
(Wiley, New York, 1966).

\bibitem{[16]}
M. Fannes,  Contribution to this volume.

\bibitem{[17]}
R. Alicki, {\it Phys. Rev.} {\bf A65}, 034104-1 (2002).

\bibitem{[18]}
S.G. Rautian and A.M. Shalagin,
{\it Kinetic problems of non-linear spectroscopy}
(North-Holland, Amsterdam, 1991). 

\bibitem{[19]}
R. Alicki and S. Kryszewski, 
{\it Completely positive Bloch-Boltzmann equations},
(physics/0202001)
 

\bibitem {[20]}
Ph. Blanchard and A. Jadczyk,
{\it Ann. der Physik} {\bf 4}, 583 (1995)


\bibitem{[21]}
R. Alicki and M. Fannes, {\it Lett. Math. Phys.} {\bf 11}, 259 (1986).


\bibitem{[22]}
R. Alicki and M. Fannes, 
{\it Commun. Math. Phys.} {\bf 108}, 353 (1987).

\bibitem{[23]}
H.J. Carmichael, 
{\it An Open Systems Approach to Quantum Optics},
(Springer, Berlin, 1993).

\bibitem{[24]}
W. Strunz, 
{\it Contribution to this volume}.


\bibitem{[25]}
R.L. Hudson and K.R. Parthasarathy, {\it Commun. Math. Phys.} {\bf 93}, 301 (1984).

\bibitem{[26]}
A.S. Holevo,  {\it J. Math. Phys.} {\bf 37}, 1812 (1996).

\bibitem{[27]}
E.B. Davies and H. Spohn,
{\it J. Stat. Phys.} {\bf 19}, 511 (1978).

\bibitem{[28]}
R. Alicki, 
{\it J. Phys.} {\bf A 12}, L103 (1979).

\bibitem{[29]}
P. Pechukas, 
{\it Phys. Rev. Lett.} {\bf 73}, 1060 (1994); {\bf 75}, 3021 (1995);
R. Alicki,
{\it ibid.} {\bf 75}, 3020 (1995). 

\bibitem{[30]}
S. Weinberg,
{\it The Quantum Theory of Fields},
(Cambridge University Press, Cambridge, 2000).


\bibitem{[31]}
A.J. Leggett et al.,
{\it Rev. Mod. Phys.} {\bf 59}, 1 (1987).

\bibitem{[32]}
D. Goderis, A. Verbeure and P. Vets, 
{\it Probability Theory and Related Fields} {\bf 82}, 527 (1989).


\bibitem{[33]}
L. Accardi et. al., in {\em Quantum Probability and Related Topics},Vol.6,
  237, (World Scientific, Singapore, 1992).


\bibitem{[34]}
E.B. Davies, 
{\it Commun. Math. Phys.} {\bf 39}, 91 (1974).


\bibitem{[35]}
V. Gorini and A. Kossakowski , 
{\it J. Math. Phys.} {\bf 17}, 1298  (1976).


\bibitem{[36]}
R. D\"umcke, 
{\it Commun. Math. Phys.} {\bf 97}, 331 (1985).


\bibitem{[37]}
G. Lindblad, G., 
{\it Commun. Math. Phys.} {\bf 40}, 147 (1975).

\bibitem{[38]}
R. Alicki,
{\it Int. J. Theor. Phys.} {\bf 16}, 351 (1977).

\bibitem{[39]}
W. Pusz and S.L. Woronowicz,
{\it Commun. Math. Phys.} {\bf 58}, 273 (1978).

\bibitem{[40]}
H. Spohn and J. Lebowitz,
{\it Adv. Chem. Phys.} {\bf 38}, 109 (1978).

\bibitem{[41]}
A.E. Allahverdyan and Th. M. Niewenhuizen,
{\it Phys. Rev.} {\bf E 64}, 056117 (2001).

\bibitem{[42]}
R.Alicki, M. Horodecki, P. Horodecki and R. Horodecki,
{\it Phys.Rev.} {\bf A} (in print 2002).

\bibitem{[43]}
E. Joos and H.D. Zeh,
{\it Z.Phys.} {\bf B 59}, 223 (1985).

\bibitem{[44]}
W.H. Zurek,
{\it Physics Today} {\bf 40}, 36 (1991)


\bibitem{[45]}
W. G. Unruh,
{\it Phys. Rev.} {\bf A 51}, 992 (1995).

\bibitem{[46]}
L. Viola and S. Lloyd,
{\it Phys. Rev.} {\bf A 58}, 2733 (1998).

\bibitem{[47]}
A.O. Caldeira and A.J. Leggett,
{\it Phys. Rev.} {\bf A 31}, 1057 (1985) 


\bibitem{[48]}
G.G Emch,
{\it Algebraic Methods in Statistical Mechanics and Quantum Field
  Theory}.
(Wiley, New York, 1972).


\bibitem{[49]}
R. Alicki,
{\it Phys. Rev.} {\bf A 40}, 4077 (1989).


\bibitem{[50]}
F. Haake,
Contribution to this volume.


\bibitem{[51]}
B. Vacchini,
{\it J. Math. Phys.} {\bf 42}, 4291 (2001); contribution to this volume. 


\bibitem{[52]}
W. A. Majewski, Contribution to this volume.

\end{references}
\end{document}